\documentclass{article}
\usepackage{spconf}

\usepackage{amsmath,amsfonts,bm}

\def\eqref#1{equation~\ref{#1}}

\def\1{\bm{1}}

\def\rmA{{\mathbf{A}}}

\def\rmR{{\mathbf{R}}}
\def\rmS{{\mathbf{S}}}

\def\ermA{{\textnormal{A}}}

\def\ermR{{\textnormal{R}}}
\def\ermS{{\textnormal{S}}}

\def\vh{{\bm{h}}}

\DeclareMathAlphabet{\mathsfit}{\encodingdefault}{\sfdefault}{m}{sl}
\SetMathAlphabet{\mathsfit}{bold}{\encodingdefault}{\sfdefault}{bx}{n}

\def\gE{{\mathcal{E}}}

\def\gG{{\mathcal{G}}}

\def\gI{{\mathcal{I}}}

\def\gN{{\mathcal{N}}}

\def\gU{{\mathcal{U}}}
\def\gV{{\mathcal{V}}}

\usepackage{algpseudocode}
\usepackage{algorithm}

\usepackage[utf8]{inputenc}
\usepackage{graphicx}
\usepackage{arydshln}
\usepackage{xcolor}
\usepackage{amsmath}
\usepackage{mathtools}
\usepackage{cite}
\usepackage{caption}
\usepackage{subcaption}
\usepackage{float}
\usepackage{times}
\graphicspath{{figures/}} 
\usepackage[utf8]{inputenc}

\begin{document}
\title{Preference and Concurrence Aware Bayesian Graph Neural Networks for Recommender Systems}
\name{Hongjian Gu\qquad Yaochen Hu\qquad Yingxue Zhang}
\address{Huawei Technologies Canada\\
         Montreal, Quebec, Canada}
\date{October 2022}

\maketitle
\begin{abstract}
    Graph-based collaborative filtering methods have prevailing performance for recommender systems since they can capture high-order information between users and items, in which the graphs are constructed from the observed user-item interactions that might miss links or contain spurious positive interactions in industrial scenarios. The Bayesian Graph Neural Network framework approaches this issue with generative models for the interaction graphs. The critical problem is to devise a proper family of graph generative models tailored to recommender systems. We propose an efficient generative model that jointly considers the preferences of users, the concurrence of items and some important graph structure information. Experiments on four popular benchmark datasets demonstrate the effectiveness of our proposed graph generative methods for recommender systems.
\end{abstract}

\section{Introduction}
Recommendation systems are playing an increasingly important role in online consumption applications in the era of the information flood. The essential task is to predict the set of items that each user is likely to interact with \cite{covington2016deep,resnick1997recommender}. To improve users' experience, collaborative filtering (CF) \cite{koren2009matrix,koren2008factorization,hofmann2004latent,he2020lightgcn} is proposed to recommend similar items to users with similar interests based on the past interactions between users and items. Later on, the performance of CF is boosted by equipping with deep-learning models \cite{guo2017deepfm,he2020lightgcn,defferrard2016convolutional,hamilton2017inductive,welling2016semi} that can automatically learn complex latent features and capture the non-linear similarity between users and items.

Recently, graph neural network (GNN) models have been extensively applied in recommender systems and achieved new state-of-the-art results \cite{hamilton2017inductive,wu2020graph,ying2018graph,he2020lightgcn,sun2019multi,kumar2022eflec}, since the user-item interactions in recommender systems can be naturally represented as bipartite graphs and the GNN can capture the high-order user-item interactions. However, as suggested in \cite{zhang2019bayesian,Sun2020AFF}, the observed graphs from the industry can contain noise information. In recommender systems, some edges may be misleading, where a user could click an item due to the attractive advertisement content rather than his/her true interest in the items, and a false negative tag might be added to some items if the user accidentally opens an app but never has a chance to check it. GNN models on the noisy graphs fail to consider the uncertainty of the observed graphs, leading to sub-optimal results.

To address the uncertainty issues of the graphs, \cite{Sun2020AFF,regol2022node} propose a new training framework on Bayesian Graph Neural Networks (BGNN). The critical module is to incorporate a graph generative model to sample ensembles of graphs from the observed one, and the GNN is applied on both the observed graph and the sampled graphs to generate a more robust graph representation. To efficiently generate similar but informative graphs, they propose the Node-Copy model, where the set of items for each user is randomly replaced with some other set of items from other users based on the similarity of the current user and the target user.

Despite the success of the Node-Copy model in the BGNN framework, this critical module has limitations, and the power of BGNN is not fully exploited. They lack the flexibility to generate novel graphs that can adapt to various scenarios. Specifically, Node-Copy can only suggest the existing combination of items from some existing users as the target item set. In the extreme case that all users have only interacted with a small range of items, the Node-Copy model can never generate a graph with the interaction between the users and the unobserved items. Besides, this model cannot be enhanced with some important prior beliefs about the properties of the observed graphs.

This work proposes a flexible graph generative model tailored to recommender systems for BGNN framework and boosts performance. Specifically, we propose that the generated graphs should share three important properties with the observed graph, namely, \emph{user preference}, \emph{item concurrence} and \emph{node degree distribution}. Pursuing a similar user preference guarantees that the generated graphs will not contain obvious false negative edges. Preserving item concurrence turns to stick the frequently co-appeared items as a whole and preserves valuable collaborative signals. Keeping similar node degree distribution will generate similar graphs structurally.

Bearing the three properties in mind, we propose an efficient iterative heuristic algorithm that independently generates the set of neighbouring items for each user. At each iteration, the probability of sampling some item is jointly determined by the user preference, the similarity between the item and the sampled item set and the item degree. Through extensive experiments on four public datasets, our proposed graph generative model consistently outperforms the Node-Copy model and other strong baselines.

\section{Preliminaries}
Recommender systems with collaborative filtering predict a set of items for each user that he/she may like based on past user-item interactions. We can treat the interactions as a graph $\gG=\{\gV, \gE\}$, where $\gV = \gU\cup\gI$ is the union of user nodes $\gU$ and item nodes $\gI$, and $\gE$ defines the set of edges where $(u, i)\in\gE$ if user $u$ has interacted with item $i$. We use $\gN(i)$ to denote the set of neighbor nodes for node $i$.
Since $\gG$ is a bipartite graph for recommender systems, we can represent the $\gG$ with the interaction matrix $\rmR\in\mathbb{R}^{|\gU|\times|\gI|}$, where $\ermR_{ui}:=1$ if $(u, i)\in\gE$ and otherwise $\ermR_{ui}:=0$. 

\cite{he2020lightgcn,sun2019multi} demonstrates that graph neural networks achieve superior performance due to the fact that graphs can capture high-order user-item interaction information.

As depicted in \cite{Sun2020AFF}, the observed graphs $\gG$ are noisy in industry, and they could miss positive edges or contain false positive edges. To solve this issue, \cite{Sun2020AFF} introduces the Bayesian graph neural networks (BGNN) framework, and they incorporate a graph generative model to generate new graphs $\hat{\gG}$ based on the observed graph $\gG$. The final representation $\vh_i$ of a node $i\in\gV$ is defined as the concatenation of the embedding from the original graph $\vh_i^{\gG}$ and the sampled graphs $\vh_i^{\hat{\gG}}$, i.e., $\vh_i = \vh_i^\gG|| \vh_i^{\hat{\gG}}$, $\forall i\in\gV$.

One of the core issues of the BGNN framework is to design a graph generative model $p(\hat{\gG}|\gG)$. \cite{Sun2020AFF,regol2022node} propose to use the node-copy model, where for each user, the connected item set is randomly replaced with the item set from other users based on some predefined user-user similarity. We argue that this approach lacks the flexibility of generating novel graphs that extract the intrinsic nature of recommender systems and thus is doomed to be sub-optimal. In this work, our goal is to propose a systematic and flexible graph generative model that captures several critical factors for recommender systems and further boosts the performance of BGNN.

\section{Methodology}

\subsection{Design Choices for Graph Generative Models}
\label{sec:design}
The core problem of designing $p(\hat{\gG}|\gG)$ is to define $p(\hat{\rmR}|\gG)$ or $p(\hat{\rmA}|\gG)$, where $\hat{\rmR}$ and $\hat{\rmA}$ are the corresponding matrix representation of $\hat{\gG}$. In order to extract the intrinsic information from recommender systems and preserve sufficient flexibility of the generative model, we consider the model
\begin{align}
    p(\rmR|\gG) = \prod_i p_i(\rmR_{i*} | \gG),
\end{align}
where $\rmR_{i*}$ denotes the $i^{th}$ row of the matrix $\rmR$. Intuitively, we model the distribution of the set of items that each user connects to, and we need to design a proper distribution $p_i(\rmR_{i*} | \gG)$ for each user $i$. Besides, we devise the following three factors that the generated graphs should be consistent with the original graphs, namely, \emph{user preference}, \emph{item concurrence} and \emph{node degrees}.

\subsubsection{User preference}
\label{sec:preference}
Although the graph generative model generates random graphs, we still want the users to connect to the items that they have potential interests in and avoid the obvious fake links to suppress noises. Inspired by the stochastic block model \cite{HOLLAND1983109,li2016scalable}, we first cluster the users and items. Then we count the number of interactions between every pair of user-item clusters in the original graph. Specifically, for clustering users, we define the pairwise user-user similarity $d(u,v)$ for $u, v\in\gU$ as
\begin{align}
    d(u, v) = \frac{\langle \rmR_{u*}, \rmR_{v*} \rangle}{|\rmR_{u*} + \rmR_{v*}|_0}, \quad \forall u, v\in\gU,
\end{align}
where $\rmR_{u*}$ denotes the $u^{th}$ row of matrix $\rmR$, $\langle \cdot,\cdot\rangle$ denotes dot product and $|\cdot|_0$ denotes $l_0$ norm. Then we adopt DB-scan \cite{schubert2017dbscan}, an efficient spectral cluster method \cite{liu2018spectral}, to cluster the users. We adopt a similar approach for items. Denotes $c_u$ as the cluster index for user $u$ and $c_i$ as the cluster index for item $i$. For each pair of user $u$ and item $i$, the user preference of user $u$ over item $i$ is related to $e(c_u, c_i)$ given by
\begin{align}
    e(c_u, c_i) = \sum_{v: c_v=c_u}\sum_{j: c_j=c_i} \ermR_{vj}. \label{eq:cluster_score}
\end{align}

\subsubsection{Item Concurrence}
In recommender systems, certain combinations of items could be frequently observed in different users' interactions. To name a few, a user who bought a TV might probably buy a TV antenna; a user who bought flower seeds is very likely to buy a flower pot; a user who bought a pen could be interested in buying a bottle of ink. In our graph generative model, we incorporate such an item concurrence signal. Specifically, we define a matrix $\rmS^\gG\in\mathbb{R}^{|\gI|\times|\gI|}$ to measure the concurrence of items on the graph $\gG$ by
\begin{align}
    \ermS^\gG_{ij} = \begin{cases}
                         \frac{\langle \rmR_{*i}, \rmR_{*j} \rangle}{|\rmR_{*i} + \rmR_{*j}|_0},  \quad & \forall i\neq j,\\
                         0, \quad &\text{otherwise}.
                     \end{cases}
\end{align}
where $\rmR_{*i}$ denotes the $i^{th}$ column of matrix $\rmR$, $\langle \cdot,\cdot\rangle$ denotes dot product and $|\cdot|_0$ denotes $l_0$ norm. We enforce that $\mathbb{E}(\rmS^{\hat{\gG}})$ should be close to $\rmS^\gG$. Although we take a similar definition for the item-item concurrent matrix $S_{ij}$ with the item-item similarity measure $d(i,j)$ defined in Sec.~\ref{sec:preference}, we argue that these are just a well-performed setting from our experiments. However, we can take different definitions for $S_{ij}$ and $d(i,j)$ for potential improvement.

\subsubsection{Node Degree}
Node degree is one of the common statistics \cite{guo2022systematic} to characterize the structure of graphs. We aim to keep a consistent structure for the generated graphs in that the node degree of each node should be similar to it from the original graph. Specifically, in the original graph $\gG$, the node degree for a node $i\in\gV$ is
\begin{align}
    d^\gG_i = \sum_j \ermA_{ij}.
\end{align}
In the sampled graph $\hat{\gG}$, we enforce that
\begin{align}
    \mathbb{E}(d^{\hat{\gG}}_i) = d^\gG_i, \quad \forall i \in \gV.
\end{align}

\subsection{Preference and Concurrence (PECO) Aware Bipartite Graph Generation}
We propose an efficient iterative heuristic algorithm to implement the sampling procedure for each user $u$ that takes the design choices in Sec.~\ref{sec:design} into account. Specifically, we need to sample a set of items for each user. To control the deviation of the sampled graph from the original graph, we initialize the set $\hat{S}_u$ of items by uniformly randomly sampling a proportion $r$ of items from its original item set, where $r\in[0, 1]$ is some tune-able hyper-parameter. Then we iteratively sample one item and add it to $\hat{S}_u$ until $|\hat{S}_u| = |\gN(u)|$. The probability of sampling item $i$ is defined as
\begin{align}
    p_u(i) \propto q_u(i) + \alpha\cdot s_{\hat{S}_u}(i), \label{eq:iter_sample_prob}
\end{align}
where $q_u(i)$ defines a preference probability, $s_{\hat{S}_u}(i)$ defines the concurrence score of the item $i$ against the sampled set $\hat{S}_u$, and $\alpha$ is some trade-off hyper-parameter controlling how much item concurrence signal we want to add in the sampled graphs. Specifically, $q_u(i)$ is a normalized value of
\begin{align}
    q_u(i) \propto \frac{e(c_u, c_i)}{|\{j| c_j=c_i \forall j\in\gI\}|}d_i^\gG,
\end{align}
where $|\{j| c_j=c_i \forall j\in\gI\}|$ is the size of cluster $c_i$, $e(c_u, c_i)$ is defined in (\ref{eq:cluster_score}) and $d_i^\gG$ is the node degree of item $i$ in graph $\gG$. $s_{\hat{S}_u}(i)= \frac{1}{|\hat{S}_u|}\sum_{j\in\hat{S}_u} \ermS_{ij}$ is the similarity measure between item $i$ and the sampled set $\hat{S}_u$. To sample a graph $\hat{\gG}=\{\hat{\gV}, \hat{gE} \}$, we initialize the graph with the same nodes as the original graph $\gG$. We sample edges for each of the users and combine all the edges as the edge set $\hat{\gE}$. Algorithm~\ref{alg:main} presents the detailed procedure.

\begin{algorithm}[t]
\caption{PECO Bipartite Graph Generation}\label{alg}
\begin{algorithmic}
\State Initialize $\hat{\gG}=\{\hat{\gV}, \hat{\gE}\}$ with $\hat{\gV}\gets \gV$ and $\hat{\gE}\gets\emptyset$
\State For $u$ in $\hat{\gU}$
    \State \hspace{\algorithmicindent} Initialize $\hat{S}_u$ with $r|\gN(u)|$ uniformly randomly sampled items from $\gN(u)$
    \State \hspace{\algorithmicindent} While $|\hat{S}_u|<|\gN(u)|$
        \State \hspace{\algorithmicindent} \hspace{\algorithmicindent} sample $i$ according to (\ref{eq:iter_sample_prob})
        \State \hspace{\algorithmicindent} \hspace{\algorithmicindent} $\hat{S}_u\gets\hat{S}_u\cup \{i\}$
        \State \hspace{\algorithmicindent} $\hat{\gE}\gets\hat{\gE}\cup\{(u, i) |\forall i \in\hat{S}_u \}$
    \State \Return $\hat{\gG}$
\end{algorithmic}
\label{alg:main}
\end{algorithm}

\section{Experiments}
\subsection{Experiment Setting}
\textbf{Datasets}. We evaluate and compare our proposed algorithms with baselines on four public datasets of various domains, sizes and connection densities: Amazon-Beauty, MovieLens-1m, yelp2018, and Amazon-CDs. Amazon-Beauty and Amazon-CDs are subsets of Amazon-review, a popular dataset for product recommendations. MovieLens-1m contains ratings for movies. The yelp dataset is a subset of Yelp's businesses, reviews, and user data. For each user, we reserve 20\% of the ground-truth interacted items as validation and testing sets respectively, and adopt the remaining 60\% data as the training set. Table~\ref{Statistics of public datasets} summarizes the
statistics of all the datasets.

\begin{table}[t]
\centering
\caption{Statistics of public datasets.}
\begin{tabular}{lllll}
\hline
Dataset &  \# Users & \# Items & \# Interacts & Density \\ \hline
Amazon-Beauty & 7068 & 3750 & 70506 & 0.299\% \\
MovieLens-1m & 6034 & 3247 & 574631 & 2.932\% \\
yelp2018 & 45919 & 45538 & 930030 & 0.044\% \\
Amazon-CDs & 43169 & 35648 & 777426 & 0.051\% \\ \hline

\end{tabular}
\label{Statistics of public datasets}
\end{table}

\textbf{Evaluation Metrics}. For all experiments, we evaluate the recommendation accuracy by recall and NDCG for the 20 top-rated items.

\textbf{Baselines}. To demonstrate the effectiveness, we compare our proposed algorithms with the following methods:
\begin{enumerate}
    \item \textit{Multi-Graph Convolution Collaborative Filtering:} Multi-GCCF \cite{sun2019multi}
    \item \textit{Baseline of graph convolutional neural network:} LightGCN \cite{he2020lightgcn}
    \item \textit{Bayesian graph neural network with node-copying graph-generative model:} Node-Copy \cite{Sun2020AFF}
\end{enumerate}
Note that LightGCN is one of the SOTA models for GNN-based recommender systems \cite{berg2017graph,wang2019neural,wu2020graph,ying2018graph,zhang2019star}. Multi-GCCF is the backbone GNN model for both the BGNN and our PECO.

\textbf{Hyper-parameters.} Table~\ref{tab:parameter} shows the important hyper-parameters we adopt for different datasets. We select those hyper-parameters based on the best validation performance.

\subsection{Discussion on the Results}
{\bf Main results.} In Table~\ref{overall_performance_comparison}, we list the performances of 4 trials with different random seeds. We can see that our proposed PECO outperforms all the other algorithms across all datasets. Specifically, both Node-Copy and PECO are based on Bayesian graph neural networks (BGNN), and they consistently perform better than their backbone model Multi-GCCF, which indicates the effectiveness of the BGNN. Besides, our PECO robustly outperforms Node-Copy, since our generative model covers a wider span of graph generative models and our proposed design choices can effectively capture the characteristics of recommender systems. Interestingly, from the hyper-parameter settings in Table~\ref{tab:parameter}, the optimal setting of $\alpha$ is different from dataset to dataset. The MovieLens-1m dataset does not benefit from the item concurrence signal while the Amazon-Beauty dataset favours a generative model with a strong item concurrence signal. Our design of the generative model is flexible to adapt to the different datasets and boost the final performance.

\begin{figure}[t]
  \includegraphics[width=\linewidth]{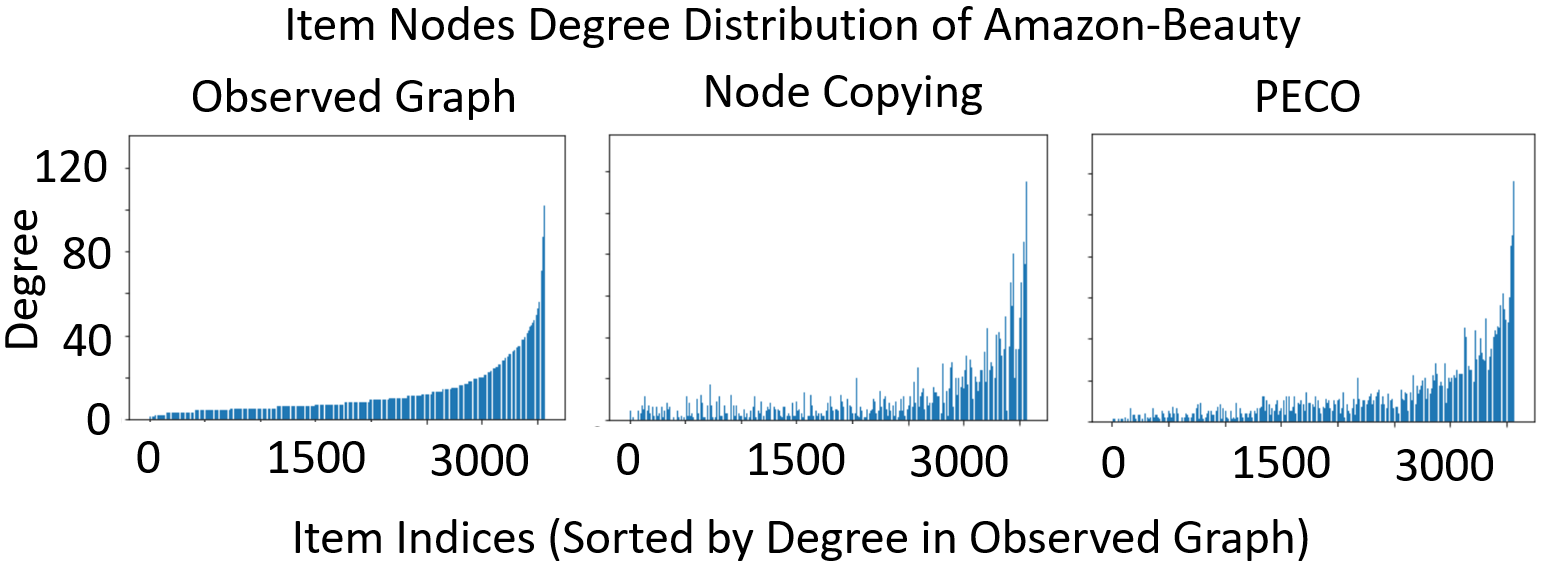}
  \caption{The item degree distribution. Item indices are sorted according to the item degree in the original graph.}
  \label{Item Degree Preservation Analysis}
\end{figure}

{\bf Statistics of sampled graphs.} We analyze how the generated graphs preserve the proposed design choices in Sec.~\ref{sec:design}. Figure~\ref{Item Degree Preservation Analysis} depicts the item degree distribution for different graphs. Note that our PECO will always result in the exact user degree distribution since we sample the same number of item nodes for each user as in the observed graph.
\begin{table}[t]
\caption{Important hyper-parameter settings for PECO.}
\begin{tabular}{llllllll}
\hline
Dataset & learning rate & \# epochs & $\alpha$ & r  \\ \hline
Amazon-Beauty & 0.01 & 500 & 1000 & 0 \\
MovieLens-1m & 0.001 & 500 & 0 & 0 \\
yelp2018 & 0.002 & 500 & 100 & 0.5 \\
Amazon-CDs & 0.0025 & 400 & 10 & 0 \\ \hline
\end{tabular}
\label{tab:parameter}
\end{table}

\begin{table}[t]
\centering
\caption{The overall performance comparison.}
\label{overall_performance_comparison}
\begin{tabular}{lllll}
\hline
Amazon-Beauty & Recall@20 & NDCG@20  \\ \hline
Multi-GCCF & 0.1194 & 0.0654 \\
LightGCN & 0.1109 & 0.0621 \\
Node-Copy & 0.1247 & 0.0701 \\ \hdashline
PECO & 0.1265 & 0.0706 \\ \hline

\hline
MovieLens-1m & Recall@20 & NDCG@20  \\ \hline
Multi-GCCF & 0.2212 & 0.2239 \\
LightGCN & 0.2238 & 0.2319 \\
Node-Copy & 0.2223 & 0.2258 \\ \hdashline
PECO & 0.2245 & 0.2280 \\ \hline

\hline
yelp2018 & Recall@20 & NDCG@20  \\ \hline
Multi-GCCF & 0.0804 & 0.0481 \\
LightGCN & 0.0717 & 0.0429 \\
Node-Copy & 0.0812 & 0.0485 \\ \hdashline
PECO & 0.0820 & 0.0489 \\ \hline

\hline
Amazon-CDs & Recall@20 & NDCG@20  \\ \hline
Multi-GCCF & 0.1135 & 0.0683 \\
LightGCN & 0.1055 & 0.0636 \\
Node-Copy & 0.1167 & 0.0705 \\ \hdashline
PECO & 0.1172 & 0.0708 \\ \hline

\end{tabular}
\end{table}

\section{Conclusion}
In this work, we propose a bipartite graph generation algorithm that exploits the preferences of users and the concurrence relationships of items. We applied it to Bayesian Graph Neural Networks replacing the node copying graph generative model. From our extensive experiments, our proposed method demonstrated consistent recommendation
accuracy improvement over the node-copying algorithm and other benchmarks for four public benchmark datasets. Potential future research directions include more sophisticated node classification algorithms and learning the trade-off parameter $\alpha$.

\bibliographystyle{IEEEbib}
\bibliography{main}

\end{document}